\documentclass[prl,showpacs,amssymb,floatfix,twocolumn]{revtex4}
\usepackage{amsmath}
\usepackage{graphicx}
\usepackage{epsfig}
\usepackage{dcolumn}
\usepackage{bm}
\usepackage{times}

\def\(({\left(}
\def\)){\right)}

\def\[[{\left[}
\def\]]{\right]}

\newcommand{\be}{\begin{equation}}
\newcommand{\ee}{\end{equation}}
\newcommand{\bea}{\begin{eqnarray}}
\newcommand{\eea}{\end{eqnarray}}

\begin{document}

\title{Hiding Quiet Solutions in Random Constraint Satisfaction
  Problems}

\author {Florent Krzakala$^{1,2}$ and Lenka Zdeborov\'a$^{2}$}
\affiliation{ $^1$ CNRS and ESPCI ParisTech, 10 rue Vauquelin, UMR
  7083 Gulliver, Paris 75000 France \\
  $^2$Theoretical Division and Center for Nonlinear Studies, Los
  Alamos National Laboratory, NM 87545 USA}

\begin{abstract}
  We study constraint satisfaction problems on the so-called {\it planted}
  random ensemble. We show that for a certain class of
  problems, e.g. graph coloring, many of the properties of the usual
  random ensemble are quantitatively identical in the planted random
  ensemble. We study the structural
  phase transitions, and the easy/hard/easy pattern in the average
  computational complexity. We also discuss the finite temperature
  phase diagram, finding a close connection with the
  liquid/glass/solid phenomenology.
\end{abstract}
  

\maketitle

Constraint satisfaction problems (CSPs) stand at the root of the
theory of computational complexity \cite{Cook71} and arise in computer
science, physics, engineering and many other fields of
science. Consider a set of $N$ discrete variables and a set of $M$
Boolean constraints; the problem consists in finding a configuration
of variables that satisfies all the constraints or in proving that no
such configuration exists.  Algorithmical approaches to 
intrinsically hard NP-complete CSPs \cite{Cook71} are one of the
biggest challenges in today's science.  Ensembles of random CSPs,
where the constraints are chosen uniformly at random from some
prescribed distribution, are being used to understand the average
computational complexity
\cite{MitchellSelman92,CheesemanKanefsky91}. Techniques from
statistical physics of glassy systems have allowed us to shed new light
on the problem and on the origin of the average algorithmical
hardness~\cite{MezardZecchina02,KrzakalaMontanari06,ZdeborovaMezard08}.

A major point in evaluating the performance of new algorithms for hard CSPs
is to be able to generate difficult instances that are guaranteed to
be satisfiable. {\it Planting} is the most standard way to do so: one
first chooses a configuration of variables and {\it then} considers
only constraints which are compatible with this planted
configuration. Many planting protocols have been introduced
\cite{BarthelHartmann02, HaanpaaJarvisalo06,LiMaZhou08}, however, the
understanding of when and why they provide a difficult instance is
still very poor compared to what is known for the purely random
ensemble \cite{MezardZecchina02, KrzakalaMontanari06,
  ZdeborovaMezard08}. This is because planting a solution changes the
properties of the ensemble. It is moreover often anticipated that the
planted solution is easier to find than a random one, as has been
indeed proven for high density of constraints
\cite{Coja-OghlanKrivelevich07,AltarelliMonasson07,Ben-ShimonVilenchik07}.
Hard instances with a known solution are also appealing to
cryptographic application as they provide good one-way functions.
Planted instances may also result from applications where only
constraints compatible with an initial state of the system are added.

In this Letter we show that for a specific, yet large, class of CSPs,
one can easily generate planted instances by hiding a {\it quiet} solution
that does not have influence on most of the characteristics of the
ensemble.  The canonical example of a CSP where a solution can be planted
in the {\it quiet} way is the graph $q$-coloring problem on which we shall
illustrate our findings about the phase diagram and the average
algorithmical hardness. The class of problems that allow a simple
quiet hiding will be discussed towards the end of the Letter.

\paragraph*{Hiding without changing ---}
The graph coloring problem consists in deciding if the $N$ vertexes
of a graph can be colored using only $q$ colors in such a way that
every two adjacent vertices have different colors. The control
parameter is the average degree of variables $c$, and we consider
the thermodynamical limit $N\to \infty$. In statistical physics, this
problem corresponds to a Potts antiferromagnet
\cite{MuletPagnani02,ZdeborovaKrzakala07}.

The way to plant a quiet solution in the graph coloring problem is actually the most
natural one: One assigns a random color with equal probability to each
of the $N$ vertices, and then constructs the graph by randomly throwing
links between vertices of different colors. Using the cavity
method \cite{MezardParisi01} we describe the
phase diagram and the structure of solutions in this planted
ensemble. In the large $N$ limit, the degree distribution in the planted graphs is
Poissonian with mean $c$, and thus they are locally tree-like
just as the standard random Erd\"{o}s-R\'{e}nyi graphs. Following the cavity approach
\cite{MezardParisi01,ZdeborovaKrzakala07} the Belief-Propagation (BP)
equations can be written. Denote $\psi^{i \to j}_s$ the
probability that the site $i$ takes color $s$ in absence of the site
$j$:
\be 
\psi^{i\to j}_s =
f(\{{\psi^{k\to i} \}}) = \frac{1}{Z^{i\to j}} {\prod_{k\in \partial
    i \setminus j} \left(1-\psi^{k\to i}_s \right) }\, , \label{BP}
\ee 
where $Z^{i\to j}$ is a normalization ensuring that $\sum_{s=1}^q
\psi^{i\to j}_s=1$. The entropy (the logarithm of the number of proper
colorings) is computed from the fixed point of eq.~(\ref{BP}) as
\bea S&=& \sum_i
\log{\Big[ \sum_{s=1}^q {\prod_{j\in \partial i} \left(1-\psi^{j\to
        i}_s \right) }\Big] }\nonumber \\ &-&
 \sum_{(ij)}\log{ \Big( 1
  - \sum_{s=1}^q \psi^{j\to i}_s \psi^{i\to j}_s \Big) }\, .\label{entropy} 
\eea
The entropy per site $s=S/N$ can thus be computed if the distribution
$P(\psi)$ over the graph is known. Assuming the absence of long range
correlations, recursive equations on this distribution can be written and
solved via the population dynamics technique \cite{MezardParisi01}. In
the planted ensemble one needs to distinguish between the sites that
were planted with different colors, we thus consider $q$ different distributions:
\bea P_{s}(\psi) &=& \sum_{k=0}^{\infty} \frac{e^{-c} c^k}{k!}
\frac{1}{ (q-1)^k} \sum_{s_1\dots,s_k}\nonumber \\ && \int
\prod_{i=1}^k \[[ P_{s_i}(\psi^i)\,{\rm d} \psi^i \]] \delta \[[ {
  \psi-f(\{{\psi^i\}}) } \]]\, , \label{BP_pl} 
\eea
where $s_i$ are all the possible colors but $s$, $s$ is taking values
$1,\dots,q$, and function $f(\cdot)$ was defined in eq.~(\ref{BP}).
The fixed point of (\ref{BP_pl}) may depend on the initial
conditions. One might initialize $P_s(\psi)$ randomly, or
in the planted solution itself, i.e., all the
elements in $P_s(\psi)$ are vectors fully oriented in the direction of
the color $s$. The dependence on initial conditions is a generic sign
for the appearance of different Gibbs states.

Before discussing further properties of the planted ensemble let
us review briefly those of the purely random ensemble.  The space of
solutions in the coloring of random graphs undergoes several
transitions as average degree $c$ is increased
\cite{KrzakalaMontanari06,ZdeborovaKrzakala07}: For low enough degree, $c<c_d$, almost all
solutions (proper colorings of the graph) belong to a single Gibbs
state and the problem can be studied using the BP approach. For $c>c_d$,
the space of solutions shatters into exponentially many different
clusters/states, each corresponding to a different Gibbs state. In this case,
a technique called one-step replica symmetry breaking (1RSB)
\cite{MezardParisi01,MezardZecchina02} is used to describe the phase
space. To focus on clusters of a given size we introduce the Parisi 1RSB parameter $m$, clusters are then weighted by their size to the power of $m$ \cite{Rivoire05,ZdeborovaKrzakala07}. For $c<c_c$ a typical solution belongs to a cluster corresponding to the value $m=1$. For $c>c_c$, although exponentially many clusters exist, a random solution will with high probability belong to
one of the few largest clusters, corresponding to $0<m<1$, while the $m=1$ clusters do not exist anymore; this is
called the condensed phase \cite{KrzakalaMontanari06}. Finally for
$c>c_s$ the last cluster disappears and no solutions exist
anymore. On top of this geometrical behavior in the space of solutions,
a remarkable phenomenon appears within the clusters themselves. In
some of them a finite fraction of variables are allowed only one
color (a phenomenon call {\it freezing)}
\cite{ZdeborovaKrzakala07,Semerjian07}. To the best of our knowledge,
no existing algorithm is able to find solutions in the frozen clusters
in polynomial time \cite{ZdeborovaKrzakala07,DallAstaRamezanpour08,ZdeborovaMezard08}, and since in a region near to the colorability threshold all clusters are frozen this provides a bound on the algorithmically hard phase.

Coming back to the planted ensemble, notice that eq.~(\ref{BP_pl}) is
nothing else but the 1RSB equation for the coloring of purely random
graphs at $m=1$ (compare e.g. with eq.~(C4) in
\cite{ZdeborovaKrzakala07}, or with the equations for the
reconstruction on trees \cite{MezardMontanari06}). It is known from
\cite{MezardMontanari06} that if (\ref{BP_pl}) is initialized in the
planted configuration then in the fixed point the distribution $P_s$
is biased towards color $s$ above the reconstruction threshold, i.e.,
for average degree $c>c_d$ \cite{MezardMontanari06}. The value $c_d$
is then a spinodal point for the existence of a {\it planted} Gibbs
state containing the planted configuration. From the equivalence of
eq.~(\ref{BP_pl}) with the 1RSB equation at $m=1$ for the purely
random ensemble it also follows that both the planted and the purely
random ensembles admit the so-called liquid solution where all
$\psi_s=1/q$. A~linear stability analysis shows that the liquid
solution is locally stable against small perturbations towards the
planted solution for $c<c_l=(q-1)^2$ \footnote{This corresponds to the
usual  local spin glass instability in the the purely random ensemble
  \cite{ZdeborovaKrzakala07}.}. Above $c_l$ the only stable fixed
point of (\ref{BP_pl}) is strongly biased towards the planted
configuration and $c_l$ is therefore a spinodal point for the liquid
state. The fact that the liquid solution is stable for $c<c_l$ also
means that the properties of the phase space are not affected by the
very existence of the planted state. This leads us to the important
conclusion ---- which we call {\it quiet} planting--- that in this
region the properties of the planted ensemble are exactly the same as
the properties of the purely random ensemble, up to the existence of
the planted state \footnote{In fact $q!$ planted states, due to the
  color-permutation symmetry.}.

For completeness let us mention that, just as in the purely random
ensemble \cite{ZdeborovaKrzakala07}, the liquid solution in the
planted ensemble decomposes further into 1RSB states for $c>c_d$. Properties of these states can be obtained by solving the 1RSB
equations. In the planted ensemble, the 1RSB equations have only one
nontrivial solution independent from the planted configuration
\footnote{It can be shown that this solution is locally stable against
  perturbations towards the planted configuration for $c<c_l$.} and it
is identical to the 1RSB solution in the purely random ensemble. Since
the liquid state is identical in the two ensembles, it is not
so surprising than the same conclusion applies to the glassy phase.

\paragraph*{Phase diagram of planted coloring ---}
\begin{figure}
\hspace{-0.5cm}
\includegraphics[width=8.8cm]{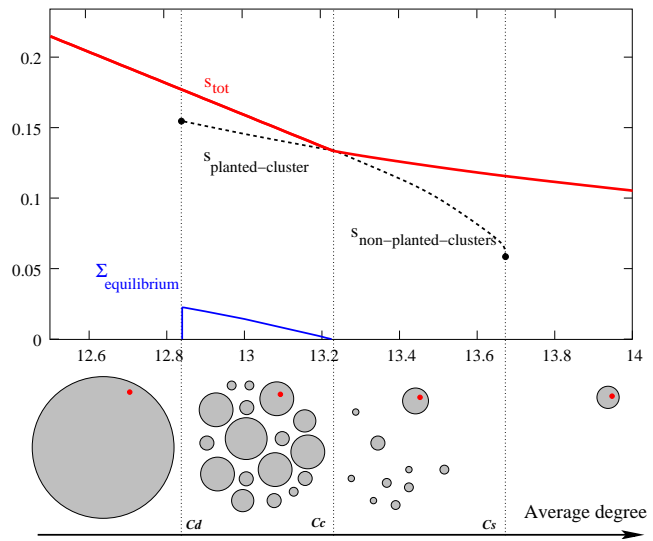}
\caption{\label{Fig:1} (color online) Phase diagram on the
  $5$-coloring on the planted ensemble. Bottom: Sketches of the
  clustering. At an average degree $c_d$ the space of
  solutions shatters into exponentially many clusters, the planted
  cluster being one of them. Beyond $c_c$ the planted cluster contains
  more solutions than all the others together. At $c_s$ the last
  non-planted cluster disappears. Top: Total entropy $s_{\rm tot}$
  with the sub-dominant part (dashed). The
  equilibrium complexity $\Sigma_{\rm equilibrium}$ (logarithm of the
  number of dominant clusters), the entropy of the non-planted
  clusters and critical degrees are taken from
  \cite{ZdeborovaKrzakala07}.}
\end{figure}
We now describe the phase diagram of the planted ensemble
(Fig.~\ref{Fig:1}). Up to the average degree $c_d$ almost all
solutions belong to one single large cluster/state of entropy $s_{\rm
  BP}=\log q+(c/2)\log (1-1/q)$.  Above~$c_d$ the space of solutions
splits into exponentially many clusters, as in the purely random
ensemble. As the planted cluster/state is described by the solution of
(\ref{BP_pl}) it has all the properties of the $m=1$ clusters from the
purely random ensemble. In particular, for $c_d<c<c_c$ the planted
cluster is one of the exponentially many equilibrium clusters and thus
for $c<c_c$ the purely random and planted ensembles of random graphs
are asymptotically equivalent.  Interestingly, this equivalence has
been rigorously proven in \cite{AchlioptasCoja-Oghlan08}, however only
up to an average degree $c_q<c_c$\footnote{With $c_q=3.83$(q=3),
  $c_q=7.81$(q=4) and $c_q\to c_c$ as $q \to \infty$.}.

For $c>c_c$ all the non-planted $m=1$ clusters disappear and the size of
the planted cluster becomes larger than the total size of the
remaining ones. A first order transition happens and the planted
state dominates the total number of solution. The entropy is
given by the fixed point of (\ref{BP_pl}) initialized in the planted
solution plugged into (\ref{entropy}). Another transition appears at
$c_s$ (the colorability threshold in the purely random ensemble)
beyond which all clusters disappears {\it except} the planted one. The
values of $c_d$, $c_c$ and $c_s$ given $q$ 
are identical to those in the purely random ensemble, and are listed in~\cite{ZdeborovaKrzakala07}.

\begin{figure}[t]
\includegraphics[width=8.6cm]{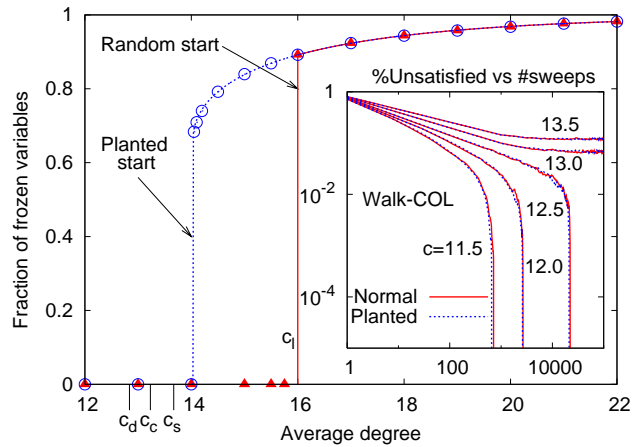}
\caption{\label{Fig:2} (color online) Fraction of variables frozen in
  their planted colors in $5$-coloring of a $N=10^5$ graph.  Data
  obtained from the BP fixed point when initialized randomly (full
  triangles) and in the planted configuration (also called the
  whitening \cite{Parisi02b}, empty circles), compared to the
  theoretical predictions \cite{ZdeborovaKrzakala07,Semerjian07} (full
  and dashed lines). For $c>c_l=16$ BP converges spontaneously to the
  planted fixed point. For $c<14.04$ \cite{Semerjian07} there are no
  frozen variables in the planted cluster. Inset: Fraction of
  monochromatic edges versus the number of sweeps of the Walk-COL
  algorithm \cite{ZdeborovaKrzakala07} in $5$-coloring of a purely
  random and a planted graph, $N=10^5$. Quiet planting does not seem
  to affect the computational hardness in the region $c<c_s$.}
\end{figure}

The properties of the planted cluster can be studied numerically on a
single graph (as was done for satisfiability in \cite{LiMaZhou08}).
We checked on many instances that the BP equations (\ref{BP})
initialized in the planted configuration converge to the liquid fixed
point for $c<c_d$, and to the planted one for $c>c_d$, while when
initialized randomly they converge to the planted fixed point only for
$c>c_l$. We have also considered the appearance of frozen variables
discussed in \cite{Semerjian07}: Fig.~\ref{Fig:2} shows the agreement
between the fraction of frozen variables obtained by applying the
whitening procedure \cite{Parisi02b} to the planted configuration on a
given graph with the theoretical prediction \cite{Semerjian07}. 

\paragraph*{Easy/Hard/Easy pattern ---}
If one does not discover the planted cluster, the
planted and the purely random graphs are indistinguishable and we thus
expect that in such a case they have comparable algorithmic difficulty. This is
indeed what we observed in experiments with several solvers. 
Fig.~\ref{Fig:2} shows results of the Walk-COL algorithm
\cite{ZdeborovaKrzakala07} on both the planted and purely random
graphs. No difference is visible, thus (unless the planted cluster
intervenes) the easy/hard pattern observed in the colorable phase $c<c_s$ is the
same in both the ensembles. It has been empirically argued this
transition is related to the freezing of clusters
\cite{ZdeborovaKrzakala07}.

On the other hand, for very large degree $c\gg c_s$ it is known that
even simple message passing algorithms find a solution near to
the planted one \cite{Coja-OghlanKrivelevich07}: therefore a second
hard/easy transition must exist. This is due to the aforementioned
linear instability at~$c_l$: Since for $c>c_l$ BP (\ref{BP}) converges
spontaneously towards the planted fixed point (as shown
in Fig.~\ref{Fig:2}) it is easy to find solutions from the
planted cluster above $c_l$ (e.g. BP decimation algorithm finds solutions in linear time). For $c<c_l$, however, without prior knowledge of the planted configuration BP converges to the uniform liquid fixed point. Applying the state of art algorithms (BP decimation,
BP reinforcement, Walk-COL, simulated annealing, etc.) to planted instances for
$c_s<c<c_l$ we were indeed not able to find solutions in polynomial
time. This suggests that the hard/easy algorithmic transition in the
planted ensemble arises exactly at $c_l$. Note at this point that
since $c_d=c_l$ for $q=3$, the planted $3$-coloring is algorithmically
easy for all degrees.

\begin{figure}[t]
  \hspace{-0.3cm}
  \includegraphics[width=8.7cm]{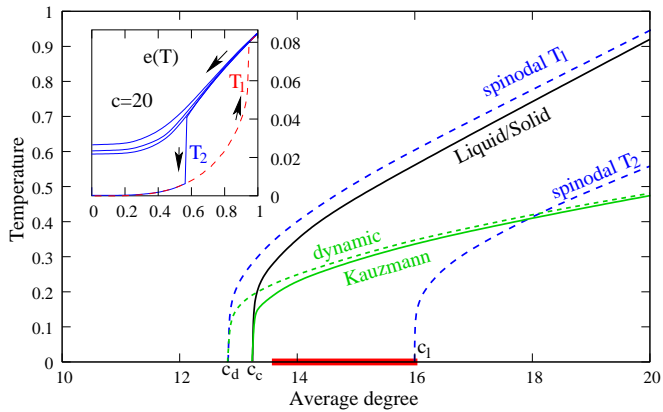}
  \caption{\label{Fig:3} (color online) Finite temperature phase
    diagram of the planted $5$-coloring. At high temperature the
    liquid is the unique Gibbs state. Below the higher spinodal line
    $T_1$ a planted (solid) Gibbs state appears, and becomes
    thermodynamically dominant at the liquid/solid transition.  The
    super-cooled liquid state is locally stable until the lower
    spinodal line $T_2$. The liquid undergoes the usual dynamical and
    Kauzmann (ideal) glass transitions (data from
    \cite{KrzakalaZdeborova07}). The thick line depicts the
    algorithmically hard region, the left boundary being the freezing transition \cite{ZdeborovaKrzakala07}.  Inset: Energy versus temperature in
    Monte-Carlo annealing with rate $\delta
    T=10^{-2},10^{-3},10^{-4}$ and $10^{-5}$ per sweep for $N=5.10^5$
    at $c=20>c_l$. Above $c_l$ a slow enough annealing undergoes a
    transition towards the planted state.}
\end{figure}

\paragraph*{Phase diagram at finite temperature ---} 
It is finally of interest to consider the properties of the problem at
finite temperature $T$, using a unit energy cost for every
monochromatic edge. The BP and the 1RSB equations
can be easily extended to this situation as e.g. in
\cite{ZdeborovaKrzakala07,KrzakalaZdeborova07}.  Note, however, that at
$T>0$ the equations for the planted ensemble do not correspond anymore
to the 1RSB equations at $m=1$, making the finite temperature problem
richer. In fact, the system behaves just as a usual mean field glass
problem, with its liquid/glass transition, where the planted state
acts as a solid-like (or crystal) phase. This
solid/planted phase exists bellow an upper spinodal temperature $T_1$
(that starts at $c_d$, see Fig.~\ref{Fig:3}), and the liquid solution
$\psi_s=1/q$ becomes unstable towards this solid state as it
encounters a spinodal point at 
\be T_2 = -1/\log{\[[\frac {c-(q-1)^2}{q-1+c}\]]}\, , \ee which starts
at $c_l$ at $T=0$ (see Fig.~\ref{Fig:3}). As usual in first order
phase transitions, the free energy of the liquid and solid state have
to be compared to draw the equilibrium phase transition line, starting
at $c_c$ at $T=0$.  As in the purely random ensemble the liquid state
undergoes a dynamical and Kauzmann glass transition
(Fig. \ref{Fig:3}). The 3-coloring is particular: the two
spinodals coincide, making the equilibrium transition of a second
order. A similar phase diagram as in Fig.~\ref{Fig:3} was found in
\cite{FranzMezard01} for the ferromagnetic p-spin model, in fact that
model is just a particular case of our quiet planting
setting. In~\cite{FranzMezard01}, however, the liquid state is always
stable and finding the ground state is polynomial since the problem
reduces to a set of linear equations.

The inset of Fig.~\ref{Fig:3} shows the behavior of Monte-Carlo
annealing for $c>c_l$ to illustrate the liquid/glass/solid
phenomenology: Upon lowering the temperature, the liquid can be
super-cooled, the solid phase avoided and a glass transition observed.
But with slow enough annealing the system transits to the solid
(planted) state at temperature $T_2$. In this case a simple simulated
annealing is able to find the ground state. If the system is
initialized in the planted solution and the temperature is increased,
the solid will melt to the liquid state at temperature $T_1$. For
connectivities $c\le c_l$ the absence of the liquid spinodal line and
the mean field nature of the model (barriers between states are
extensive) makes algorithms based on local dynamics unable to find the
planted cluster. This gives a physical interpretation behind the
hard/easy transition at $c_l$.

\paragraph*{Conclusion ---}
We have discussed the graph coloring problem on the planted ensemble
and showed that quantitative results, explicitly checked via numerical
experiments, can be readily deduced from what is known in the purely
random ensemble. In particular we showed that for $c<c_c$ the planted
ensemble is asymptotically equivalent to the purely random one.  Many
works have established the easiness of the planted ensemble at very
large
\cite{Coja-OghlanKrivelevich07,AltarelliMonasson07,Ben-ShimonVilenchik07}
and very small average degree. We bridged the gap and showed that
while the easy/hard transition in the planted ensemble is similar to
the one in the usual random ensemble, the hard/easy transition
coincides with a local instability of the liquid phase at
$c_l=(q-1)^2$. This leaves a large region of very hard problems with a
known hidden solution. We also showed how to create mean field
``glass'' models with a solid-like/planted state.

Let us finish by discussing the (large) class of CSPs where the quiet
planting is possible and where
our phenomenology readily apply. The crucial property that we used when stating that the
natural planting does not change much of the structural properties was
the uniformity of the BP fixed point in the purely random ensemble
(e.g. in coloring $\psi=(1/q,\dots,1/q)$). Many other CSPs actually
share this property, e.g. all the problems without disordered
interactions on random regular graphs, the hyper-graph bicoloring
\cite{DallAstaRamezanpour08}, or the balanced locked problems of
\cite{ZdeborovaMezard08}. These last ones are particularly appealing
as hard satisfiable benchmarks. The random satisfiability problem,
however, is a canonical example where the fixed point of the BP equation is not uniform and where our results do not
apply. It would be interesting to generalize our approach to plant quiet solutions in such cases.

\vspace{-0.5cm}

\bibliography{myentries}

\begin{thebibliography}{25}
\expandafter\ifx\csname natexlab\endcsname\relax\def\natexlab#1{#1}\fi
\expandafter\ifx\csname bibnamefont\endcsname\relax
  \def\bibnamefont#1{#1}\fi
\expandafter\ifx\csname bibfnamefont\endcsname\relax
  \def\bibfnamefont#1{#1}\fi
\expandafter\ifx\csname citenamefont\endcsname\relax
  \def\citenamefont#1{#1}\fi
\expandafter\ifx\csname url\endcsname\relax
  \def\url#1{\texttt{#1}}\fi
\expandafter\ifx\csname urlprefix\endcsname\relax\def\urlprefix{URL }\fi
\providecommand{\bibinfo}[2]{#2}
\providecommand{\eprint}[2][]{\url{#2}}

\bibitem[{\citenamefont{Cook}(1971)}]{Cook71}
\bibinfo{author}{\bibfnamefont{S.~A.} \bibnamefont{Cook}}, in
  \emph{\bibinfo{booktitle}{Proc. 3rd STOC}} (\bibinfo{publisher}{ACM},
  \bibinfo{address}{NY, USA}, \bibinfo{year}{1971}).

\bibitem[{\citenamefont{Mitchell et~al.}(1992)\citenamefont{Mitchell, Selman,
  and Levesque}}]{MitchellSelman92}
\bibinfo{author}{\bibfnamefont{D.~G.} \bibnamefont{Mitchell}},
  \bibinfo{author}{\bibfnamefont{B.}~\bibnamefont{Selman}}, \bibnamefont{and}
  \bibinfo{author}{\bibfnamefont{H.~J.} \bibnamefont{Levesque}}, in
  \emph{\bibinfo{booktitle}{Proc. 10th AAAI}} (\bibinfo{publisher}{AAAI Press},
  \bibinfo{address}{Menlo Park, California}, \bibinfo{year}{1992}).

\bibitem[{\citenamefont{Cheeseman et~al.}(1991)\citenamefont{Cheeseman,
  Kanefsky, and Taylor}}]{CheesemanKanefsky91}
\bibinfo{author}{\bibfnamefont{P.}~\bibnamefont{Cheeseman}},
  \bibinfo{author}{\bibfnamefont{B.}~\bibnamefont{Kanefsky}}, \bibnamefont{and}
  \bibinfo{author}{\bibfnamefont{W.~M.} \bibnamefont{Taylor}}, in
  \emph{\bibinfo{booktitle}{Proc. 12th IJCAI}} (\bibinfo{publisher}{Morgan
  Kaufmann}, \bibinfo{address}{San Mateo, CA, USA}, \bibinfo{year}{1991}).

\bibitem[{\citenamefont{M{\'e}zard and Zecchina}(2002)}]{MezardZecchina02}
\bibinfo{author}{\bibfnamefont{M.}~\bibnamefont{M{\'e}zard}} \bibnamefont{and}
  \bibinfo{author}{\bibfnamefont{R.}~\bibnamefont{Zecchina}},
  \bibinfo{journal}{Phys. Rev. E} \textbf{\bibinfo{volume}{66}},
  \bibinfo{pages}{056126} (\bibinfo{year}{2002}).

\bibitem[{\citenamefont{{Zdeborov{\'a}} and
      {M{\'e}zard}}(2008{\natexlab{a}})}]{ZdeborovaMezard08}
  \bibinfo{author}{\bibfnamefont{L.}~\bibnamefont{{Zdeborov{\'a}}}}
  \bibnamefont{and}
  \bibinfo{author}{\bibfnamefont{M.}~\bibnamefont{{M{\'e}zard}}},
  \bibinfo{journal}{Phys. Rev. Lett.} \textbf{\bibinfo{volume}{101}},
  \bibinfo{pages}{078702} (\bibinfo{year}{2008}{\natexlab{a}}); {and}
  \bibinfo{journal}{J. Stat. Mech.}
   \bibinfo{pages}{P12004}
  (\bibinfo{year}{2008}{\natexlab{b}}).


\bibitem[{\citenamefont{Krzakala et~al.}(2007)\citenamefont{Krzakala,
  Montanari, Ricci-Tersenghi, Semerjian, and
  Zdeborov{\'a}}}]{KrzakalaMontanari06}
\bibinfo{author}{\bibfnamefont{F.}~\bibnamefont{Krzakala}} {\it et al},
  \bibinfo{journal}{Proc. Natl. Acad. Sci. U.S.A}
  \textbf{\bibinfo{volume}{104}}, \bibinfo{pages}{10318}
  (\bibinfo{year}{2007}).

\bibitem[{\citenamefont{Barthel et~al.}(2002)\citenamefont{Barthel, Hartmann,
  Leone, Ricci-Tersenghi, Weigt, and Zecchina}}]{BarthelHartmann02}
\bibinfo{author}{\bibfnamefont{W.}~\bibnamefont{Barthel}} {\it et al},
  \bibinfo{journal}{Phys. Rev. Lett.} \textbf{\bibinfo{volume}{88}},
  \bibinfo{pages}{188701} (\bibinfo{year}{2002}).

\bibitem[{\citenamefont{Haanp\"{a}\"{a}
  et~al.}(2006)\citenamefont{Haanp\"{a}\"{a}, J\"{a}rvisalo, Kaski, and
  Niemel\"{a}}}]{HaanpaaJarvisalo06}
\bibinfo{author}{\bibfnamefont{H.}~\bibnamefont{Haanp\"{a}\"{a}}} {\it et al.},
  \bibinfo{journal}{JSAT} \textbf{\bibinfo{volume}{2}}, \bibinfo{pages}{27}
  (\bibinfo{year}{2006}).
\bibinfo{author}{\bibfnamefont{D.}~\bibnamefont{Achlioptas}} {\it et al}, in
  \emph{\bibinfo{booktitle}{Proc. of AAAI-2000, Austin, Texas, USA}}
  (\bibinfo{year}{2000}). \bibinfo{author}{\bibfnamefont{D.}~\bibnamefont{Achlioptas}},
  \bibinfo{author}{\bibfnamefont{H.}~\bibnamefont{Jia}}, \bibnamefont{and}
  \bibinfo{author}{\bibfnamefont{C.}~\bibnamefont{Moore}},
  \bibinfo{journal}{JAIR}
  \textbf{\bibinfo{volume}{24}}, \bibinfo{pages}{623} (\bibinfo{year}{2005}).
\bibinfo{author}{\bibfnamefont{C.}~\bibnamefont{Moore}},
  \bibinfo{author}{\bibfnamefont{H.}~\bibnamefont{Jia}}, \bibnamefont{and}
  \bibinfo{author}{\bibfnamefont{D.}~\bibnamefont{Strain}}, in
  \emph{\bibinfo{booktitle}{Proc. AAAI}} (\bibinfo{year}{2005}).


\bibitem[{\citenamefont{Zhou}(2008)}]{LiMaZhou08}
\bibinfo{author}{\bibfnamefont{K. Li, H. Ma, H. Zhou}} 
  (\bibinfo{year}{2008}), \bibinfo{note}{arXiv:0809.4332}.


\bibitem[{\citenamefont{F.~Altarelli and Zamponi}(2007)}]{AltarelliMonasson07}
\bibinfo{author}{\bibfnamefont{R.~Monasson} \bibnamefont{F.~Altarelli}}
  \bibnamefont{and} \bibinfo{author}{\bibfnamefont{F.}~\bibnamefont{Zamponi}},
  \bibinfo{journal}{J. Phys. A: Math. Theor.} \textbf{\bibinfo{volume}{40}},
  \bibinfo{pages}{867} (\bibinfo{year}{2007}).

\bibitem[{\citenamefont{Ben-Shimon and
  Vilenchik}(2007)}]{Ben-ShimonVilenchik07}
\bibinfo{author}{\bibfnamefont{S.}~\bibnamefont{Ben-Shimon}} \bibnamefont{and}
  \bibinfo{author}{\bibfnamefont{D.}~\bibnamefont{Vilenchik}}, in
  \emph{\bibinfo{booktitle}{Proc. of the 13th International Conference on
  Analysis of Algorithms, DMTCS}} (\bibinfo{year}{2007}).

\bibitem[{\citenamefont{Coja-Oghlan
      et~al.}(2007)\citenamefont{Coja-Oghlan, Krivelevich, and
      Vilenchik}}]{Coja-OghlanKrivelevich07}
  \bibinfo{author}{\bibfnamefont{A.}~\bibnamefont{Coja-Oghlan}},
  \bibinfo{author}{\bibfnamefont{M.}~\bibnamefont{Krivelevich}},
  \bibnamefont{and}
  \bibinfo{author}{\bibfnamefont{D.}~\bibnamefont{Vilenchik}}, in
  \emph{\bibinfo{booktitle}{Proc.  STACS 24, LNCS 4393}}
  (\bibinfo{year}{2007}).

\bibitem[{\citenamefont{Mulet et~al.}(2002)\citenamefont{Mulet, Pagnani, Weigt,
  and Zecchina}}]{MuletPagnani02}
\bibinfo{author}{\bibfnamefont{R.}~\bibnamefont{Mulet}} {\it et al},
  \bibinfo{journal}{Phys. Rev. Lett.} \textbf{\bibinfo{volume}{89}},
  \bibinfo{pages}{268701} (\bibinfo{year}{2002}).

\bibitem[{\citenamefont{Zdeborov{\'a} and
  Krzakala}(2007)}]{ZdeborovaKrzakala07}
\bibinfo{author}{\bibfnamefont{L.}~\bibnamefont{Zdeborov{\'a}}}
  \bibnamefont{and} \bibinfo{author}{\bibfnamefont{F.}~\bibnamefont{Krzakala}},
  \bibinfo{journal}{Phys. Rev. E} \textbf{\bibinfo{volume}{76}},
  \bibinfo{pages}{031131} (\bibinfo{year}{2007}).

\bibitem[{\citenamefont{M{\'e}zard and Parisi}(2001)}]{MezardParisi01}
\bibinfo{author}{\bibfnamefont{M.}~\bibnamefont{M{\'e}zard}} \bibnamefont{and}
  \bibinfo{author}{\bibfnamefont{G.}~\bibnamefont{Parisi}},
  \bibinfo{journal}{Eur. Phys. J. B} \textbf{\bibinfo{volume}{20}},
  \bibinfo{pages}{217} (\bibinfo{year}{2001}).



\bibitem{Rivoire05}
\bibnamefont{M. M\'ezard, M. Palassini and O. Rivoire,}
  \bibinfo{journal}{Phys. Rev. Lett.} \textbf{\bibinfo{volume}{95}},
  \bibinfo{pages}{200202} (\bibinfo{year}{2005}).


\bibitem[{\citenamefont{Semerjian}(2008)}]{Semerjian07}
\bibinfo{author}{\bibfnamefont{G.}~\bibnamefont{Semerjian}},
  \bibinfo{journal}{J. Stat. Phys.} \textbf{\bibinfo{volume}{130}},
  \bibinfo{pages}{251} (\bibinfo{year}{2008}).



\bibitem[{\citenamefont{M\'ezard and Montanari}(2006)}]{MezardMontanari06}
\bibinfo{author}{\bibfnamefont{M.}~\bibnamefont{M\'ezard}} \bibnamefont{and}
  \bibinfo{author}{\bibfnamefont{A.}~\bibnamefont{Montanari}},
  \bibinfo{journal}{J. Stat. Phys.} \textbf{\bibinfo{volume}{124}},
  \bibinfo{pages}{1317} (\bibinfo{year}{2006}).


\bibitem[{\citenamefont{Parisi}(2002)}]{Parisi02b}
\bibinfo{author}{\bibfnamefont{G.}~\bibnamefont{Parisi}}
  (\bibinfo{year}{2002}), \bibinfo{note}{arXiv:cs.CC/0212047}.

\bibitem[{\citenamefont{Krzakala and Zdeborov\'a}(2008)}]{KrzakalaZdeborova07}
\bibinfo{author}{\bibfnamefont{F.}~\bibnamefont{Krzakala}} \bibnamefont{and}
  \bibinfo{author}{\bibfnamefont{L.}~\bibnamefont{Zdeborov\'a}},
  \bibinfo{journal}{EPL} \textbf{\bibinfo{volume}{81}},
  \bibinfo{pages}{57005} (\bibinfo{year}{2008}).

\bibitem[{\citenamefont{Franz et~al.}(2001)\citenamefont{Franz, M{\'e}zard,
  Ricci-Tersenghi, Weigt, and Zecchina}}]{FranzMezard01}
\bibinfo{author}{\bibfnamefont{S.}~\bibnamefont{Franz}} {\it et al},
  \bibinfo{journal}{Europhys. Lett.} \textbf{\bibinfo{volume}{55}},
  \bibinfo{pages}{465} (\bibinfo{year}{2001}).


\bibitem[{\citenamefont{Achlioptas}(2008)}]{AchlioptasCoja-Oghlan08}
\bibinfo{author}{\bibfnamefont{D. Achlioptas, A. Coja-Oghlan}} 
  (\bibinfo{year}{2008}), \bibinfo{note}{arXiv:0803.2122}.


\bibitem[{\citenamefont{Dall'Asta}(2008)}]{DallAstaRamezanpour08}
\bibinfo{author}{\bibfnamefont{L. Dall'Asta {\it et al.}}},
\bibinfo{pages}{031118} \bibinfo{journal}{Phys. Rev. E} \textbf{\bibinfo{volume}{77}} (\bibinfo{year}{2008}).


\end{thebibliography}

\end{document}